%% file: narr.tex
\documentclass[onecolumn]{article}
\usepackage{narr}

\usepackage{amssymb,epsfig,graphicx,amssymb,subeqnarray, multirow, pstricks,colortbl, bm}
\usepackage{units,tabularx,array,booktabs,booktabs,algorithm, algorithmic}
\usepackage[cmex10]{amsmath}
\usepackage[tight,footnotesize]{subfigure}  
\include{macpap2}

\usepackage{paralist}
\usepackage{hyperref}
\usepackage{soul}
\usepackage{float}
\usepackage[font={small}]{caption}

\definecolor{Gray}{gray}{0.9}
\newcolumntype{g}{>{\columncolor{Gray}}c}
\begin{document}

\title{A hierarchical narrative framework for OCD}

\author{P.J. Moore${}^\dag$ }

\address{
\dag Mathematical Institute, Oxford University.\\
}

\maketitle

\abstract
This paper gives an explanatory framework for obsessive-compulsive disorder (OCD) based on a generative model of cognition.  The framework is constructed using the new concept of a `formal narrative' which is a sequence of cognitive states inferred from sense data.  First we propose that human cognition uses a hierarchy of narratives to predict changes in the natural and social environment.   Each layer in the hierarchy represents a distinct `view of the world', but it also contributes to a global unitary perspective. Second, the generative models used for cognitive inference can create new narratives from those states already experienced by an individual.  We hypothesise that when a threat is recognised, narratives are generated as a cognitive model of possible threat scenarios.   Using this framework, we suggest that OCD arises from a dysfunction in sub-surface levels of inference while the global unitary perspective remains intact.  The failure of inference is felt as the external world being `not just right', and its automatic correction by the perceptual system is experienced as compulsion.  Ordering and symmetry obsessions are the effects of the perceptual system trying to achieve precise inference.  Checking behaviour arises because the security system attempts to finesse inference as part of its protection behaviour.  Similarly, fear of harm and distressing thoughts occur because the failure of inference results in an indistinct view of the past or the future.  A wide variety of symptoms in OCD is thus explained by a single dysfunction.
\endabstract
% Note: need to add names to the bare references - or change the citation format.
% keywords: mental time travel,


%____________________________________________________________________________________________________
%
%   S E C T I O N
%_____________________________________________________________________________________________________
% Should update neuropsychiatric review using Kersten 2004.
% Cite enright1994 paper which suggests OCD could be related to schizophrenia.
% Cite my thesis.
%
% Patients commonly exhibit personality traits for example: `tender conscience' \cite{reed1985},  perfectionism \cite{frost}, low self-directedness \cite{cruz}.
%\cite[p182-214]{antony02}
% Could use Jakes book to expand
\section{Symptoms and empirical observations}
The symptoms of OCD are obsessions and compulsions.  Obsessions are recurrent and persistent thoughts, impulses, or images which are experienced as intrusive and inappropriate, and they cause marked anxiety or distress \cite{dsmtr}.  Compulsions are repetitive behaviours or mental acts that the person feels driven to perform.  Some examples of symptoms are: repeatedly checking gas taps or washing hands, fearing having knocked someone down while driving, and feeling an impulse to shout obscenities during a church service \cite[p196]{miller1988}.   More generally the focus of obsessions can be aggression, contamination and symmetry among others, while compulsions are often centred on checking, ordering, cleaning and hoarding \cite{dsmtr}.  The categories are related: for example obsessions about aggression, religious or sexual themes are associated with checking compulsions \cite{summerfeldt1999}\cite{leckman97}.  Some patients exhibit common cognitive traits or beliefs: 1) Responsibility and threat estimation, 2) Perfectionism and intolerance for uncertainty, and 3) Importance and control of thoughts \cite{occwg03_1}\cite{occwg03_2}, while other patients do not exhibit dysfunctional beliefs \cite{antony02}\cite{taylor06}.  Some patients believe that the intrusive thoughts can influence events in the world, a phenomenon known as `thought--action fusion' \cite{shafran1996}\cite{shafran2004}. 
\nind
In general neuropsychological investigations of OCD have not given a consistent picture of cognitive deficits.  A meta analysis of 113 studies \cite{abramovitch2013} by Abramovitch \etal reduced performance in people with OCD compared with healthy individuals across most neuropsychological domains.  Specifically, those with OCD were found to score consistently and significantly worse than controls on non-verbal memory tasks measured by the Rey–-Osterrieth Complex Figure Test.  Neuroimaging studies of OCD have found differences between patients with OCD and controls in the orbital gyrus and the head of the caudate nucleus \cite{saxena2001}\cite{whiteside2004}.  A review of evidence from both neuroimaging and neuropsychological studies is given in \cite{menzies2008}.
\nind
A review of treatments for OCD,  is provided by Ponniah \etal \cite{ponniah2013}.   This study found that exposure and response prevention (ERP) and cognitive-behavioral therapy (CBT) were efficacious and specific for OCD.  ERP and CBT have comparable efficacy (for example, \cite{whittal2005} \cite{fisher2005}) with both having recovery rates of approximately 25\% \cite{fisher2005}.

% Could also cite Milner Beech and Walker cired in security motivation paper, and the bean jar tests.

%____________________________________________________________________________________________________
%
%   S E C T I O N
%_____________________________________________________________________________________________________
\section{Theories}
Current theoretical approaches to OCD emphasise the negative appraisal of intrusive thoughts, where such appraisals are engendered by dysfunctional beliefs.  This cognitive-behavioural account of OCD begins by noting that intrusive thoughts and images, similar in content to clinical obsessions, occur generally in the population \cite{julien2007}\cite{gibbs1996}\cite{rachman1978}. The hypothesis is that in OCD such intrusions develop into obsessions when they are appraised as personally important, highly unacceptable or immoral, or as posing a threat for which the individual is personally responsible \cite{abramowitz2006}\cite{abramowitz2007}.  Compulsions arise from an attempt to remove intrusions and prevent their harmful consequences, and these actions serve to increase the frequency of intrusions by acting as a reminder of their content.  This account of OCD was originated by McFall and Wollersheim \cite{mcfall1979}, Rachman \cite{rachman1997}  and Salkovskis \cite{salkovskis1985}.  Beliefs, appraisals and symptoms are known to be associated \cite{abramowitz2006}\cite{abramowitz2007}, and this association has been adduced as evidence for cognitive models of obsessive-compulsive disorder \cite{abramowitz2007}. Some objections to the approach were articulated by Jakes \cite{jakes1989a}\cite{salkovskis1989}\cite{jakes1989b}, and a overview of criticisms within a wider context was given by Jakes in \cite{jakes}. A more recent critique of the significance of dysfunctional appraisals and the cognitive-behavioural account is given by Cougle and Lee \cite{cougle2014}.  Some researchers have identified dimensions of harm avoidance and incompleteness as more fundamental motives that contribute to compulsive behavior \cite{summerfeldt2004}, where incompleteness is defined as an internal state of imperfection or `not just right' experience \cite{coles2003}.  
\nind
Other researchers have emphasised the importance of biological factors: a brief theoretical overview of biological and other models is provided in \cite{abramowitz2009}.  Wise and Rapoport \cite{wise1989}\cite{rapoport1990} suggested that the disorder arises from a dysfunction of the basal ganglia.   OCD can occur in childhood, associated with streptococcal infections, as part of the paediatric autoimmune neuropsychiatric disorders associated with streptococcal infections (PANDAS) syndrome \cite{swedo2002}\cite{leonard2001}.  The related hypothesis is that OCD (and tic disorders) arise from post-streptococcal immunity.  Another, less widely used theoretical approach is a mathematical (complex systems) approach to model the pathophysiology of the disorder \cite{rolls}.  A neuropsychological model, relevant to the current study,  explains OCD as a disturbance of security motivation \cite{szechtman2004}.  In that account the symptoms of OCD stem from an inability to generate a `feeling of knowing' that normally terminates the expression of a security motivation system. Some criticisms of this model were given in \cite{taylor2005} and a response was given in \cite{woody2005}.  The context of the current study is that there are difficulties with the dominant theoretical approach \cite{jakes}\cite{cougle2014} and there remains as yet no definitive account of OCD cf. \cite[p88]{taylor06} \cite[p165]{jakes}.  

%Jakes\cite{jakes} reviewed the approaches to OCD, and covers behavioural/learning, personality theories derived from Pavlov's work, Janet's approach, cognitive style, biological and psychodynamic approaches. 
% The quality of such explanations should be judged by whether they lead to successful new predictions which other theories cannot explain \cite[p154]{jakes}. 

%____________________________________________________________________________________________________
%
%   S E C T I O N
%_____________________________________________________________________________________________________
\section{A hierarchical narrative framework}
\subsection{Introduction}
% Note: the history of perception: Friston, O'Callaghan and Hohwy could be expanded a little.
The symptoms of OCD suggest no obvious common cause and the disorder is heterogeneous, possibly with different subtypes \cite{mckay2004}.  Psychological models for OCD do not go far beyond a description of the symptoms or traits of the patient \cite{jakes} and so they have weak explanatory power.  Biological models explain observed abnormalities in, for example imaging data, but they do not provide the level of description needed to explain symptoms.  In explaining psychopathology, the different levels of explanation -- formal electrophysical models at the neural level, and qualitative psychological explanations of symptoms -- have tended to be disjoint.  There is some work that attempts to close this `explanatory gap', in particular the \emph{Bayesian brain} hypothesis \cite{knill}\cite{fristonhist} which has been applied to both neuroscientific observations \cite{fristoncort} and symptoms of mental illness \cite{fletcher}\cite{fristonhal}.  Under this approach the brain is hypothesised to maintain probabilistic models of the environment and update the models using Bayesian inference \cite{knill}.  More specifically, the brain minimises the discrepancy between sensory input and the predictions made by an internal model, and in this way it implements Bayesian inference.  The explanatory power of prediction error minimisation for perception is explored in depth in the text by Hohwy \cite{hohwy}.  The Bayesian brain concept is usually traced to  Helmholtz's theory of perception \cite{helm}\cite{southall1925} in which stimuli are seen as insufficient to generate percepts without prior information enabling unconscious inference.  O'Callaghan describes the development of perceptual theory from Helmholtz to a contemporary understanding \cite[p78]{frankish2012}, and Friston provides a short history of the Bayesian Brain idea \cite{fristonhist} from a neuroscientific point of view.  
\nind
Two relevant applications of predictive coding to psychopathology are as follows.  Fletcher and Frith \cite{fletcher} use a hierarchical model of brain function to explain the positive symptoms of schizophrenia -- hallucinations and delusions. In a hierarchical model, prediction errors at a low level are passed up the hierarchy until they are resolved or `explained away' by a higher level.  In schizophrenia there is a failure in inference which leads to improper integration of new evidence and a resulting prediction error.  Corlett \etal \cite{corlett} focus on delusions and again propose a predictive coding model to help explain them.  Following Helmholtz, they define the brain as an inference machine, and they understand delusions as false inferences.  Predictive learning and prediction error are general mechanisms of brain function and they relate them to neurotransmitter signalling, and to experimental evidence of the effect of, for example, NMDA receptor antagonists.  Their hypothesis is that aberrant prediction error leads to aberrant learning and this in turn leads to delusions and perceptual aberrations. The approach taken in this paper also uses a hierarchical model of cognition and its dysfunction, in this case to explain the symptoms of OCD.

%____________________________________________________________________________________________________
%
%   S E C T I O N
%_____________________________________________________________________________________________________
\subsection{Hierarchical inference}
\begin{figure}[!htp]
\centering
\includegraphics[trim = 0mm 0mm 0mm 0mm, width=12cm]{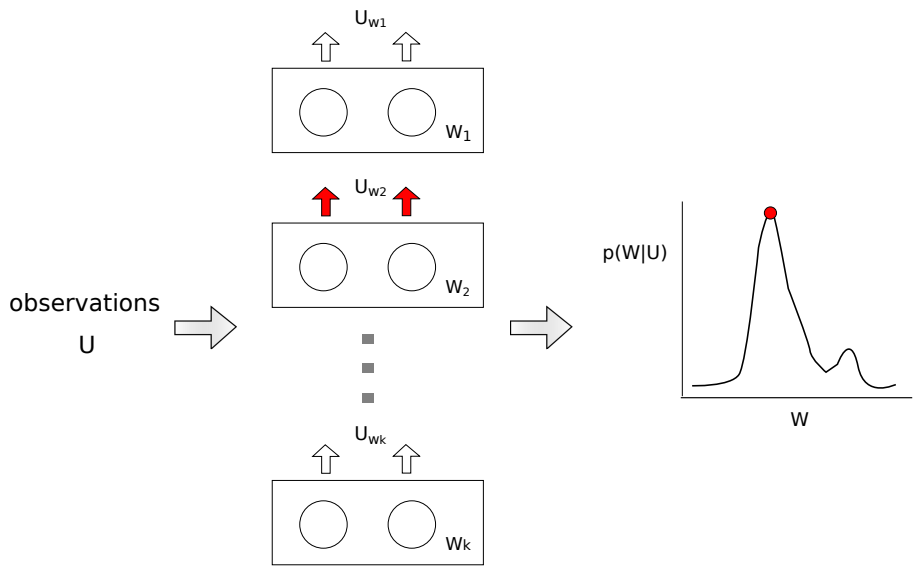} %left bottom right top
\caption[Perception as cognitive resonance]{Perception as cognitive resonance.  The rectangular boxes represent internal state models  $W_1$ \dots $W_k$ each of which can generate simulated observations $U_{W1}$ \dots $U_{Wk}$.   Perception is the process of inferring the model that best explains current observations $U$ by using Bayes' rule to find $p(W|U)$ from $p(U|W)$.  In the figure, $W_2$ is shown as resonating strongly with the observations. The process results in the graph $p(W|U)$ which has a peak at $W_2$ (right hand side).  A familiar pattern will exhibit resonance, while closely related patterns also resonate to some extent.  For example, an infant will recognise any human face, but will show particular affinity for the mother's face.}
\label{fig:perception}
\end{figure}

We first explain hierarchical Bayesian inference as a model for human cognition. An important function for any animal is to know its environment by using observations from its senses.   We define an \emph{external state} $X$ as a cause or condition in the environment that fully determines the \emph{observations} $U$.  We define an \emph{internal state} $W$ as the cognitive representation of an external state based on the observations.  The task for the cognitive apparatus is to find the distribution of internal states conditioned on observations, $p(W|U)$ in order to model the external state. 
\nind
Figure \ref{fig:perception} illustrates the process.  The probability of each state $W_1$ \dots $W_k$ individually generating the observations $U$ is determined.  The process of inference is one of selecting the state $\hat{W}$ that is most likely to have generated the observations $U$.  The result is shown on the right hand side of the figure as a graph of $p(W|U)$ against the internal states $W$.  For a known pattern generated by a familiar state, the graph will show resonance for the corresponding model.  The internal states $W$ might be visual features such as edges, or higher level percepts such as a human face.  
%____________________________________________________________________________________________________
%
%   F I G U R E
%____________________________________________________________________________________________________

\begin{figure}[!htp]
\centering
\includegraphics[trim = 0mm 0mm 0mm 0mm, width=8cm]{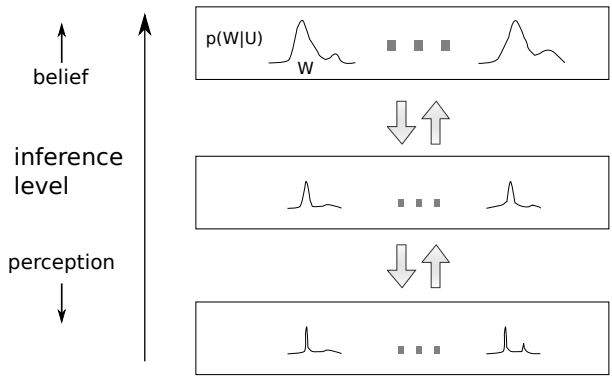} %left bottom right top
\caption[Cognition as multi-level inference of states]{A pictorial representation of cognition as multi-level inference of inferred states.   The graphs represent the conditional distribution $p(W|U)$ which has a peak at the most likely state $\hat{W}$.  Inferences from the lowest level may consist of features, such as edges in the visual scene, which become observations for the next level in the hierarchy. High level percepts needed for beliefs, thoughts and imagination are represented at the top of the hierarchy.  The perception--belief spectrum was suggested by Corlett \cite{corlett} to be relevant to delusions and hallucinations in schizophrenia.} 
\label{fig:hierarchy}
\end{figure}

% source: Inkscape: figsource/nar_perception
Higher level inference is accomplished by using a hierarchy in which states inferred by one level are used as observations for the next. Figure \ref{fig:hierarchy} shows the hierarchy of inference levels. Observations arrive from sensory neurons as spike trains which are processed into features, such as edges in a visual scene.  The features themselves become observations for a second level of inference which in turn uses them to infer states that are meaningful to the next level up.  So higher levels attempt to predict or `explain away' lower levels, so that ultimately the sense data is explained by the internal cognitive model.
% high levels predict lower levels, not sensory data  	

%____________________________________________________________________________________________________
%
%   F I G U R E
%____________________________________________________________________________________________________

% source: Inkscape: figsource/nar_hierarchy

%____________________________________________________________________________________________________
%
%   S E C T I O N
%_____________________________________________________________________________________________________

\subsection{Formal narratives}
The generative models used for inference can also generate artificial observations which have the same distribution as those already experienced.  We illustrate the process by using a generative model of language to create English sentences.  The probability of a sequence of words $W_k$ can be expanded as,
\begin{align}
p(W_k) &= p(w_1)p(w_2|w_1)p(w_3|w_1,w_2) \dots p(w_k|w_1 \dots w_{k-1}) 
\intertext{We can approximate the terms in the expansion by limiting the history length to give a bigram word model, which assumes that the probability of each word is influenced only by its predecessor,}
p(W_k) &\approx p(w_1)p(w_2|w_1)p(w_3|w_2) \dots p(w_k|w_{k-1}) 
\end{align}
% Since the joint distribution of inputs and outputs are modelled, we can then generate a new sequence from the model.  
The model then comprises the frequencies of starting words and pairs of words found in the training data. To create a new sentence, we take a starting word $w_1$ and choose the next word randomly according to its distribution in the model, and continue until the desired length is reached.  Shannon \cite[p7]{shannon} gives an example of an artificial sentence derived from such a model of word sequences,
\\ \\
\texttt{\phantom{THE} THE HEAD AND IN FRONTAL ATTACK ON AN ENGLISH WRITER THAT THE CHARACTER OF THIS POINT IS 
\\ \phantom{THE} THEREFORE ANOTHER METHOD FOR THE LETTERS THAT THE TIME OF WHO EVER TOLD THE PROBLEM FOR 
\\ \phantom{THE} AN UNEXPECTED.}
\\ \\
\comment{In the last example the first word \texttt{THE} is likely to be followed by a noun or adjective, in this case \texttt{HEAD}, and it in turn  determines the distribution of the next word.} The sentence as a whole approximates the distribution of the natural word order.  If a bigram model is used, the generated text is word salad.  Longer n-grams can be used, in which case the model comprises longer sequences of words. In this case the generated text becomes much more recognisably like English in its construction, but it usually has no coherent meaning.
\nind

%____________________________________________________________________________________________________
%
%   S E C T I O N
%_____________________________________________________________________________________________________

\begin{figure}[htpb]
\centering
\includegraphics[trim = 0mm 0mm 0mm 0mm, width=7cm]{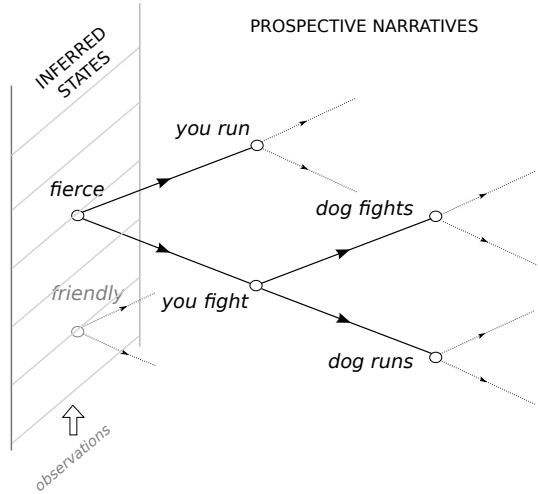} %left bottom right top
\caption[Example of prospective narratives]{Example of prospective narratives generated from a model of states.  The internal states $W_{dog-fierce}$ and $W_{dog-friendly}$ are inferred from observations using an existing model of a dog and its behaviour.  The states are those that best `explain away' or predict the observations, which are themselves inferred states from lower in the cognitive hierarchy.  A network of prospective narratives is generated, whose probabilities depend on past experience.  }%The prospective narratives shown here are generated at the time of seeing the dog, and they are updated as the observations change.}
\label{fig:narratives_prosp}
\end{figure}

We consider the same process of generation, but instead of words, we use the cognitive states $W$ as generated elements.  We call a sequence of generated states a \emph{narrative}.   To illustrate the point we use an example of a threatening scenario faced by an individual.  When someone encounters a large dog in the street, they quickly make an assessment of whether it is likely to pose a danger, and then act accordingly.  We denote the internal state representing a fierce dog as $W_{dog-fierce}$, and an internal state representing the individual running away as $W_{you-run}$, and so on.  Some prospective narratives representing this fight--or--flight scenario are,
\newline\newline
\indent\indent $N_1$ = \{$W_{dog-fierce}, \quad W_{you-run}$\} \\ %, $p(N_1)$ =  $p(W_{dog-fierce}) \, p(W_{you-run})$ \\
\indent\indent $N_2$ = \{$W_{dog-fierce}, \quad  W_{you-fight}, \quad  W_{dog-fights}$\} \\
\indent\indent $N_3$ = \{$W_{dog-fierce}, \quad  W_{you-fight}, \quad  W_{dog-runs}$\} \\
\indent\indent $N_4$ = \texttt{$\dots$} 
\newline\newline
The probability of each narrative $p(N_k) = p(W_1) p(W_2|W_1) p(W_3|W_1,W_2) \dots p(W_j|W_1,W_2 \dots W_{j-1}) $ where $j$ is the number of states in $N_k$.  In this simple model the narratives cover all possibilities, so $\sum_k p(N_k) = 1$.  The scenario is illustrated in Figure \ref{fig:narratives_prosp}.
% source: Inkscape: figsource/nar_narratives_prosp

\subsubsection{Hierarchies of narratives}
\begin{figure}[!htp]
\centering
\includegraphics[trim = 0mm 0mm 0mm 0mm, width=8cm]{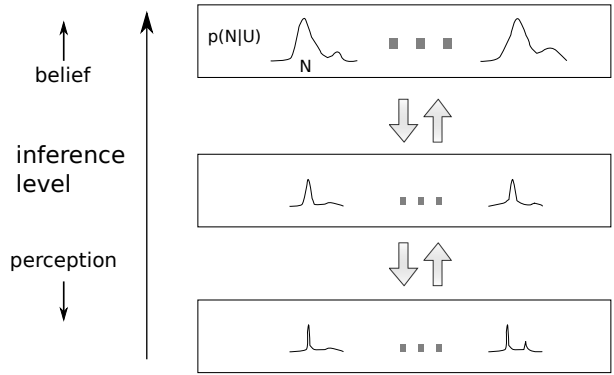} %left bottom right top
\caption[A hierarchical narrative model]{A representation of cognition as multi-level inference of narratives.   The graphs represent the conditional distribution $p(N|U)$ which has a peak at the most likely narrative $\hat{N}$.  This model is a generalisation of the one shown in Figure \ref{fig:hierarchy} because a narrative is either a single inferred state or a sequence of inferred states.   At the lowest levels the states are comprised of features derived from the different sensory modes. The conditional distribution of states at lower levels is shown as more peaked than that at higher levels, reflecting the expansion of possible inferences as we move up the cognitive hierarchy.} 
\label{fig:hierarchy_narr}
\end{figure}

The states chosen for the example represent a cognitive state in linguistic form.  For example $W_{dog-fierce}$ represents the cognitive state inferred by an individual when they see a fierce dog.  Since the example uses recognisable states, it can be presented as a narrative in the more colloquial sense.  So the formal narrative $N_3$ can be expressed retrospectively as, \emph{``I saw a fierce dog, I fought it and it ran away''}.
\nind
The substates from which the state $W_{dog-fierce}$ is inferred can not so easily be expressed in a natural language, but there is evidence that more primitive perceptions are also the outcome of a generative process.  For example, the visual hallucinations that occur in Charles Bonnet Syndrome have been proposed as evidence for a generative model of vision \cite{reichert2013}.  This observation suggests that formal narratives could occur at levels of inference below those that can be expressed in a natural language.  In Figure \ref{fig:hierarchy_narr} we generalise the model in Figure \ref{fig:hierarchy} to allow inference from narratives, where a narrative is either a single inferred state or a sequence of inferred states.  
% The example just given was chosen to illustrate the concept of a formal narrative as a sequence of inferred states. 
% There is no direct access to these states, so they are an approximation based on introspection and observed behaviour.  
%____________________________________________________________________________________________________
%
%   F I G U R E
%____________________________________________________________________________________________________

% source: Inkscape: figsource/nar_hierarchy

The model has some intuitive appeal because applying inference to narratives is common.   For example if someone hears $N_3$ as a retrospective narrative they will question if the speaker really did fight off a dog. But the model also implies that inference of narratives occurs at lower levels of cognition suggesting that the human perspective is composed of a hierarchy of layers, each of which contributes to the individual's perspective or `view of the world'.  Support for a layered model of cognition is provided by instances when individuals do not believe what they see, for example with optical illusions.  In these cases inference at a high level overrides inferences made by a subordinate level, so the individual has the experience of perception without its consequent belief.  The potential lack of coherence between levels is also relevant to OCD, where the individual recognizes that the obsessional thoughts, impulses, or images are a product of his or her own mind \cite{dsmtr}\cite{foa1995}.
% More prosaic examples are those of optical illusions or conjuring tricks.  
\subsubsection{Unitary perception and multiple inference}
\begin{figure}[htpb]
\centering
\includegraphics[trim = 0mm 0mm 0mm 0mm, width=3cm]{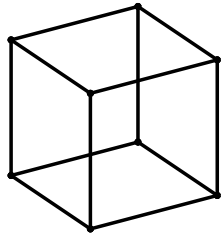} %left bottom right top
\caption[Necker Cube]{Necker Cube illustrating the unitary nature of visual perception.  Most individuals can interpret the cube in two different ways and switch between the views at will.  It is not possible to perceive both interpretations simultaneously, and the imagined solid is always coherent with the perception. The perceptual switch can be made by seeing the vertex, fourth in height on the page, as either in front of or behind the face that it intersects on the page. }
\label{fig:necker}
\end{figure}

However in normal cognition, the usual \emph{experience} of perception is unitary, and we perceive just one reality.  For example the Necker Cube, shown in Figure \ref{fig:necker} can be interpreted as a solid in two different orientations.  By fixing the perception, for example by viewing the fourth highest vertex on the page as being at the back, a solid of a given orientation is visualised. Conversely, by imagining the cube as in the alternative orientation, the perceptual view is fixed.  In either case, the imagined solid is coherent with the perception, and it is seen in only one orientation at a given time.
% Another example of coherence, in this case between different sensory modes, is the McGurk effect, in which visual input determines the perceived sound \cite{mcgurk1976}.  
%____________________________________________________________________________________________________
%
%   F I G U R E
%____________________________________________________________________________________________________

% source: necker.m
The unitary nature of perception and consciousness has a long philosophical lineage: Bayne gives an introduction in  \cite{bayneschol}, and a text of his recent work is \cite{bayne2010}.  But although the experience of perception is unitary, the inference on which it relies must entertain multiple scenarios. The Necker Cube is an ambiguous visual scene which admits two interpretations each of which is consistent with the lines on the page, and this ambiguity is processed by the cognitive apparatus.  Further, a partially obscured scene admits many interpretations because it is an under-determined problem: we have to formulate multiple hypotheses for what might be buried under the ground or hidden behind a bush.  The hierarchy of states model in Figure \ref{fig:hierarchy} fulfils this purpose by computing the most likely internal states given the sensory input\footnote{With the Necker cube, the Bayes posterior representing the configuration has two, equal maxima.}. For spatiotemporal modelling the hierarchy of narratives shown in Figure \ref{fig:hierarchy_narr} is needed.

\subsubsection{Mental time travel}
An animal derives a selective advantage from successfully interpreting and predicting changes in its natural environment: it is more likely to reproduce if it can anticipate the threat from a hidden predator, or the reward from a buried cache of food. The natural environment presents different levels of cognitive challenge.  The physical world and flora are passive, and their behaviour is relatively easy to learn.  Animals, especially predators, are harder to predict, but their threats can often be learned.  Other humans pose the most complex challenge because they are agents with the same cognitive apparatus as the individual.  The attribution of mental states to others is called  `theory of mind' or `mentalising', and it usually develops in early childhood.  Theoretical accounts of theory of mind fall into two classes: 1) the individual learns through hypothesis and experiment, in the same way that they learn about the behaviour of inanimate objects,  2) the individual employs their own cognitive apparatus to simulate another person's behaviour.   An explanation of these accounts and some of the cognitive challenges posed by simulation are given by Mitchell \cite{mitchell2009}.  A related theory is that social cognition is based on the recognition that others are `like me' \cite{meltzoff2007}, and there is some evidential support for this idea provided by Gardner \etal \cite{gardner2015}.   
\nind
The faculty of `mental time travel', a term coined by  Suddendorf and Corballis, \cite{suddendorf1997}, is also related to simulation.  It allows humans to mentally project themselves backwards in time to re-live, or forwards to pre-live, events.  The emergence of mental time travel in evolution is hypothesised to be a crucial step towards human current success \cite{suddendorf2007}.  It is suggested that mental time travel is enabled by a special conscious state called `chronesthesia', a hypothesis that has some experimental support \cite{nyberg2010}.
\nind
Helmholtz realised that the senses cannot even resolve ambiguous scenes without combining evidence with prior information \cite{kersten2004}\cite{helm}, which is a computationally difficult problem.  Modelling behaviour is harder still: it involves not just the inference of scenes, but a prediction of how they might evolve.  A solution is the constant formulation of counterfactuals and the simulation of possible scenarios.  So here we apply the simulation model not just to mentalising, but more generally for mental time travel.
Under the predictive coding model, an individual predicts their sensory data and minimises the error between that prediction and the real input.  Error minimisation serves for perception of current events, but what about prospection, which has no input against which to minimise error?  We propose that the generative models used for perception create a distribution of narratives which are constantly updated as sensory data is received.
% The Necker Cube, unitary perception section was here.
%Since chronesthesia is a sense of the past or the future it must entertain multiple scenarios.  In contrast, p
%____________________________________________________________________________________________________
%
%   S E C T I O N
%_____________________________________________________________________________________________________
%\subsubsection{Narratives as a model for mental time travel}
\nind
So a hypothesis is as follows, \emph{Narratives, or sequences of cognitive states, are created from a generative model for the purposes of mental time travel.}   Since narratives are generated from percepts and unconscious inferences as shown in Figure \ref{fig:narratives_prosp}, chronesthesia can be understood as unitary perception extended into a manifold past and future.  Just as a movie passes through time using a series of 2-dimensional scenes, a single prospective narrative can run percepts forward in time, generating for each step a \emph{prospective scene}.  Chronesthesia is the conscious awareness of examining such scenes, which might be aversive, for example being bitten by a dog or rewarding, like winning the national lottery. \comment{Similarly retrospective narratives can be used to generate scenarios that may have taken place in the past. } However, whereas conscious perception is usually veridical\footnote{Sometimes percepts do not correspond with reality, for example in the case of hallucinations or perceptual illusions.}, prospective narratives are mostly counterfactual. 
\nind
Mental time travel is based on the manifold inferences from sense data rather than on the most likely inference from that data, which is experienced as perception.  This facet of the model is relevant to `fear of harm' in OCD, in which highly unlikely scenarios are generated from seemingly false inferences.

%If imagined scenarios are believed to be real, then the individual is experiencing a delusion.  Those who score highly on positive schizotypy reported a greater feeling of mental time travel and re-living or ‘pre-living’ imagined events \cite{winfield2010}.  In OCD, imagined scenarios are not usually believed to be real but they are felt to be real.  The account presented in this paper attempts to use the concept of a narrative to explain OCD.
% The following section to be retained, but perhaps moved somewhere else.
%\nind
%The process of mental time travel seems likely to be related to default functional activity in the brain: a review of the default system is given in  \cite{morcom2007} and a critique in \cite{raichle2007}.  The default system is thought to be engaged by self-projection, including thinking about the future, episodic remembering and theory of mind \cite{buckner2007}.  
% could also include abnormalities of DMN in sz.

%____________________________________________________________________________________________________
%
%   S E C T I O N
%_____________________________________________________________________________________________________

\subsubsection{The security motivational system}
\label{sec:sms}
The risk of an adverse event is the subject of some symptoms of OCD, such as checking and fear of harm.   Szechtman and Woody \cite{szechtman2004} hypothesise that a dysfunction in the security motivational system (SMS) underlies such symptoms, specifically an inability to generate the normal `feeling of knowing' that normally signals task completion.   The security motivational system refers to a `set of biologically based (hardwired), species-typical behaviours directed toward protection from danger of self and others' \cite[p113]{szechtman2004}.  In this paper we use the term to refer to the cognitive function that invokes those behaviours.  We suggest that the SMS simulates potential behaviour by creating a collection of narratives which is used to guide the response. 
\nind
So a secondary hypothesis is as follows, \emph{In response to a threat stimulus, a collection of narratives relating to that threat is generated.} The threat stimulus is anything that has previously been associated with peril, and the narratives are constructed scenarios that involve a threat to the individual.  For example, on hearing a rustling of leaves, the individual might construct scenarios of a predator waiting to attack.  Such narratives would involve the highest levels of cognition, particularly if the threat is social.  But the response has to be fast, suggesting that the `view' of lower levels of cognition, using simpler inferred states, is important.
\nind
That humans should create threat scenarios is not surprising, when we consider threats at a corporate level.  Organisational security is routinely tested using penetration tests which use unlikely scenarios to determine how attackers can gain access to sites or data.  Significantly, these scenarios include threats which arise from within the organisation either through action or inaction.  The point here is not that the individual security mechanisms are necessarily similar to corporate security mechanisms but that in both cases the threats are open and underdetermined. So at the individual level, the construction of threats is not pathological, but a normal cognitive function.  However, within the human population there is likely to be some variation in the `creativity' of the narratives.  Highly creative narratives, although unrealistic, confer some protection in unpredictable circumstances, while uncreative narratives make for quick identification. At the extremes, we might expect there to be some dysfunction.

% \nind
% The generation of threat scenarios satifies the requirement for predicting hitherto unseen actions made by an individual.  But creative prediction must be balanced with speed of response.  Animals manage threats by binding an immediate action to a threat marker.  Humans retain this kind of response, but can direct it using higher level cognition.  

% On learning of a severe threat from for example, a small red insect, we can react immediately on seeing the insect.  In this case the narrative might be as follows,
% \newline\newline
% \indent\indent $N_1$ = \{$W_{red-insect}, \quad  W_{inaction}, \quad  W_{insect-bite}$\} \\
% \newline
% The scenario predicted by narrative $N_1$ is threatening, so the state $W_{red-insect}$ is registered as a threat marker.  The cognitive system will expressly try to `create' a red insect from the visual features.  Any approximate match, for example a substate from which of $W_{ref-insect}$ is inferred will also constitute a threat.

\section{A hierarchical narrative account of OCD}
% [this needs amending] The hypothesis for the dysfunction in OCD is that \emph{the internal states for self--generated actions have an abnormally high probability.}  The high probability leads to narratives being more likely to incorporate such states than is normal, and for those narratives which include them to have an abnormally high probability.  The justification for the hypothesis proposal is given later.  First we apply it to explain symptoms of OCD.
The notion of generated threat scenarios, including threats from within, speaks to `fear of harm' obsessions in OCD.  However it explains part of normal functioning rather than the dysfunction that leads to pathological obsessions.  The challenge for any account of OCD is to propose a dysfunction that explains symptoms and which is supported by other evidence.  For example, why are obsessions in OCD recurrent in nature, and why do they lead to marked distress \cite{dsmtr}?
\nind
OCD covers a broad range of symptoms and there is no single set of features in common and peculiar to all instances of the disorder \cite[p26]{jakes}.  Rather than confronting the problem of definition, the approach taken here is to explain specific behaviours and symptoms that are found in some individuals with OCD.  We begin by examining the results from Gillan \etalns's laboratory study of avoidance behaviour in OCD and control participants  \cite{gillan2014}.  Gillan \etalns's  study is important both for its results and for its method, which provides an experimental model for some kinds of compulsion.  So we apply the hierarchical narrative account first to Gillan \etalns's study then to specific symptoms of OCD.  %the following features of OCD: checking, intrusive thoughts and fear of harm, ordering and symmetry obsessions, and cognitive traits.
% see http://www.ocduk.org/types-ocd
%
% Perhaps include a discussion of resonance and dissonance, with peak and smiley diagrams.
%
% Note: case of guy with OCD - HIV obsession - make a note of multiple views idea - not sub-conscious.
% %____________________________________________________________________________________________________
% %
% %   S E C T I O N
% %_____________________________________________________________________________________________________
% 
% \section{Substates and action}

%____________________________________________________________________________________________________
%
%   S E C T I O N
%_____________________________________________________________________________________________________
\subsection{Gillan \etal -- Excessive avoidance habits}
Gillan \etal \cite{gillan2014} tested one possibility inspired by the cognitive-behavioural account of OCD.  That is, that excessive behavioral repetition in OCD  is driven by a failure to learn about safety.  They compared a group of 25 individuals having OCD with a matched control group.  Participants were asked to avoid an electric shock by responding to a warning stimulus presented on the screen.  Two electrodes were used, one connected to each wrist and each delivering a shock at random intervals.  The shocks were preceded by a warning stimulus whose color denoted either the right or left wrist as the target.  To avoid the shock, participants had to respond within 750ms using a corresponding left or right pedal switch.  The experimenters examined the effect of stimulus devaluation on goal--directed learning by disconnecting one of the electrodes in full view of the participant.  The task design consisted of four stages: a training session, a first devaluation test, an extended training session and a final devaluation test.  
\nind
In the first devaluation test, both the OCD group and the control group behaved in a similar way: they responded more to the valued stimulus, which represented a real shock than to the devalued stimulus, which did not.  So both OCD patients and control participants were capable of learning about safety.  After extended training, the OCD group showed greater avoidance of the devalued stimulus compared with the control group, but there were no measurable differences in contingency knowledge, explicit threat appraisal, or physiological arousal between the groups.  Those who responded to the devalued stimulus were asked to account for their action -- why did they continue to respond to the stimulus corresponding to the disconnected electrode?  The main categories of response were threat beliefs, `I thought it could still shock me', and accidental responses, `I lost concentration'.  The authors concluded that these findings supported an account of OCD involving habit formation.  How might Gillan \etalns's results be explained using the hierarchical model?  We consider three questions: 

%____________________________________________________________________________________________________
%
%   S E C T I O N
%_____________________________________________________________________________________________________
\paragraph{1) Why do some participants continue to respond to a devalued stimulus?\\}
During the experiment the security motivational system in all participants generates prospective narratives which represent the scenario in which a shock occurs. This results in an understanding, or view, that a shock will occur if the participant does not press the correct pedal when the electrode is connected and a stimulus appears.  When the electrode is disconnected the SMS continues to generate narratives, and the participant examines these narratives in order to direct their action.  Those who responded to the devalued stimulus were unable to form a clear view of the prospective situation, at the lower levels of cognitive inference\footnote{The reason some participants could not form a clear view is explained later in the section on the nature of the dysfunction.}.  If Bayesian learning fails, heuristics may act as a fall back strategy \cite{oreilly2012}, and there is evidence that in a rapidly changing environment, people act according to the last choice that they made \cite{summerfield2011}.  So the participants' action is simply based on a heuristic of selecting the last action that they performed in similar circumstances.  
%Since a heuristic is closely related to a habit, this explanation extends that in Gillan \etalns's paper.
%But the action can also be seen as an attempt to finesse the inference in order to improve their perspective, as discussed later in the section on the nature of the dysfunction.

%____________________________________________________________________________________________________
%
%   S E C T I O N
%_____________________________________________________________________________________________________
\paragraph{2) Why is some participants' understanding not consistent with their behaviour ?\\}
In the experiment, the participants are aware of the usual conduct of modern psychological experiments, the function of electrical circuits and so on.  These external conditions are coded by narratives at the highest level of cognition in both the OCD and control groups.   There are also narratives at a lower level of cognition which make use of simpler internal states representing the stimulus and the shock.  Both levels of narrative contribute to the individual's conscious view of the situation and to their evaluation of any threat. There is a failure of inference at lower levels of cognition so the individual resorts to a heuristic.  However the individual's conscious view of the situation is unitary and it is derived from the most likely inference from all cognitive layers.  This highest level of inference remains intact and results in a veridical perspective: the participants who responded to the devalued stimulus were not deluded.   Instead they found their behaviour to be inconsistent with what they felt to be their subjective view and understanding of the situation.
%____________________________________________________________________________________________________
%
%   S E C T I O N
%_____________________________________________________________________________________________________
\paragraph{3) How do we account for the participants' post-hoc explanations for their behaviour?\\}
Gillan \etal \cite{gillan2014} note that `In situations of cognitive dissonance, where behavior contradicts belief, humans are known to alter beliefs to match behavior \cite{festinger1962}. Within this framework, irrational obsessive thoughts in OCD might function to resolve the internal conflict arising from experiencing an otherwise nonsensical urge to avoid.'  Resolving the internal conflict between action and understanding necessitates creating a unitary inference to explain one's own behaviour.  Clearly this is a difficult task for the participants since they are in effect being asked to explain OCD.  According to the hierarchical narrative model, the participants' perception is internally inconsistent, so they can not present a simple answer to explain their behaviour.  Situations promoting cognitive dissonance are not uncommon: for example playing the national lottery, or the avoidance behaviour associated with rare diseases are both cases in point.  Most participants reported either subjective threat beliefs or accidental slips and these explanations are similar to given by others in situations promoting cognitive dissonance in that they focus on the threat (or reward) or on accidental events rather than on exploring the inconsistency.  

%In terms of the hierarchical narrative framework, perception is unitary and it is robust to failures of inference at low levels of perception. 
%Significantly, though, one participant in the OCD group gave the reason for responding to the devalued stimulus as a `sense of satisfaction, doing something thoroughly, conscientiously, to perfection' (Claire Gillan, personal communication).  This (admittedly selected) report suggests that the participant recognised that they were trying to gain some coherence in their cognitive model through their action.

%____________________________________________________________________________________________________
%
%   S E C T I O N
%_____________________________________________________________________________________________________
\subsubsection{The nature of the dysfunction} 
We suggest that the dysfunction in OCD is a failure of inference at sub-surface levels of the cognitive hierarchy.  When an individual has inadequate information to perform inference about a threat, they act both to improve that inference and to respond to the threat.  For example a driver slows down on approaching a traffic junction both to avoid a collision and to see the junction more clearly\footnote{The driver's behaviour can be seen as a manifestation of active inference, a unified account of both action and perception motivated by the free energy principle \cite{friston2010}\cite{fristonfree}.}. The system for monitoring threats, the SMS, usually operates silently but in OCD attention returns repeatedly to a non-existent threat.  The reason is that lower levels of inference are failing so that the individual is constantly taking action to correct their perspective and to manage a potential threat.   The urge to respond is compelling: consider the reaction of a driver whose vision suddenly becomes obscured.  This account explains why the compulsion to act is strong and why attention keeps returning to the perceived threat.  It also explains why individuals with OCD do not usually believe that a real threat exists: global inference is robust to failures of inference occurring at lower levels, just as it is robust to deficiencies in the sensory data resulting from an occluded scene.  So the individual's overall judgement is broadly unimpaired.
\nind
Among the human population there is likely to be some variation in the `creativity', in effect the number, of narratives generated for threat simulation.  If there is a propensity towards a broad distribution of $p(N|U)$  at low levels of cognition, then the posterior distribution becomes less peaked and inference becomes more uncertain.  So we locate the dysfunction in the generative process, which in turn causes a failure in inference.
\subsection{Explaining symptoms of OCD} 
This section applies the hierarchical narrative models to symptoms of OCD: checking, intrusive thoughts and fear of harm, and ordering and symmetry obsessions.  
%____________________________________________________________________________________________________
%
%   S E C T I O N
%_____________________________________________________________________________________________________
\subsubsection{Checking\\}
\vspace{-4mm}
An example of checking in OCD cited by de Silva \cite[p196]{miller1988} is of someone who repeatedly checks that a gas tap is turned off.  The explanation is broadly the same as that for the results of Gillan \etalns's experiment given in the last section.  The individual with OCD has a correct understanding that the gas tap is off even though there is a failure of sub-surface levels of inference.  The individual's SMS relies on this lower level inference, which does not provide a clear `view' of the state of the gas tap.  As a result the individual has to act both to improve the inference and for self-protection.  Since inference about the external world is failing, the individual resorts to their last action under those conditions, and checks the state of the gas tap. However, the act of checking does not rectify the failure in inference so the SMS does not terminate its response to the perceived threat.
% , and this can be related approximately to Gillan \etalns's experiment, where the stimulus is the gas tap, the threat is a gas leak, and the action taken is checking
%____________________________________________________________________________________________________
%
%   S E C T I O N
%_____________________________________________________________________________________________________
\subsubsection{Fear of causing harm \\}
\vspace{-4mm}
An example of the fear of causing harm is a case reported by de Silva \cite[p196]{miller1988} of an individual thinking, without justification, that he had knocked someone down with his car.  In this case and in the case of checking there is a feeling of uncertainty about the true state of affairs -- whether the gas tap is off, and whether an accident has really occurred -- accompanied by a recognition that the obsessional thoughts are a product of his or her own mind \cite{dsmtr}\cite{foa1995}.  The central difference between the examples is that in the case of checking, the threat stimulus is present to the individual whereas with fear of harm, the threat stimulus is no longer present.  In both cases, the individual is generating retrospective narratives to examine the scene in memory, and prospective narratives to assess the threat.  
\nind
The recognition that an accident has not happened shows that the individual's overall perception is functioning correctly and that he is not deluded. In both healthy individuals and those with OCD, the SMS generates narratives and these correctly explain the likely cause: the individual perhaps experiences a bump in the road and correctly explains the cause as, for example, a pothole.   In individuals with OCD, a failure of inference at sub-surface levels of the cognitive hierarchy results in an unclear `view' of the scene.  The SMS acts both to protect the individual from a potential threat and to correct the failure of inference.  So the individual is forced to finesse their perspective, but since the dysfunction is internal their action does not terminate the SMS.
%____________________________________________________________________________________________________
%
%   S E C T I O N
%_____________________________________________________________________________________________________
\subsubsection{Intrusive thoughts\\}
\vspace{-4mm}
Hudak \etal \cite{hudak2012} report a case of intrusive thoughts associated with postpartum OCD.  A woman was afraid to be alone with her 9 week old son because of her terrifying thoughts of stuffing the baby into a microwave oven.  She also reported thoughts about stuffing her husband into a microwave oven which she found even more frightening because she realised that this was physically impossible.  The explanation for these thoughts has the same basis as for the examples of checking and fear of harm given above, but with some additions.  The SMS constantly creates prospective narratives simulating threat scenarios, in this case involving a microwave, but the dysfunction of sub-surface inference results in an unclear `view' of the generated scenarios.  Without a clear view of a threatening situation, the SMS signals a strong urge to act to correct the perception to ensure that the child is protected.  The SMS usually operates silently, and without the individual being aware of the simulated scenarios, but the action brings the process to the individual's attention.
\nind
This account identifies the dysfunction with the evaluation of thoughts rather than with their content, so it does not explain the bizarre nature of the obsession where woman thinks of stuffing her husband in the microwave.  To explain the aberrant content, we use the more precise description of the dysfunction as an unduly `creative' generative process\footnote{Increased `creativity' in the generative process may arise from the need to protect the infant, in addition to any dysfunction.} In this case the inference at higher levels will struggle to interpret their meaning, so leading to bizarre or absurd inferences.  However the global inference which informs unitary perception is intact with the result that the individual is not deluded.  

%However, as in the earlier examples of checking and fear of harm, any feasible protective action has no effect on the dysfunction, and the SMS does not terminate.  Infeasible actions, such as disposing of the microwave, presumably would terminate the specific obsession, but the SMS would continue to operate on other threat stimuli.

%The explanation for these thoughts has the same basis as for the examples of checking and fear of harm given above.  In this case, the `creative' aspect of the SMS mentioned in the above section on \hyperref[sec:sms]{the security motivational system} is also important.  The natural and social environment is so complex that the threats facing any animal can not be precisely determined, so threats are created by simulation in response to a threat stimulus.  In this respect the process is similar to penetration testing in which corporate security is stressed using simulated threat scenarios.  Within individuals the simulation process is part of normal cognitive functioning and it operates at a level below which social or other restrictions on its content apply.  It is usually silent, and operates without the individual being aware of the simulated scenarios.

%a propensity towards a broad distribution of $p(N|U)$  at low levels of cognition

%____________________________________________________________________________________________________
%
%   S E C T I O N
%_____________________________________________________________________________________________________
\subsubsection{Ordering and symmetry obsessions\\}
\begin{figure}[tp]
\centering
\includegraphics[trim = 0mm 0mm 0mm 0mm, width=5cm]{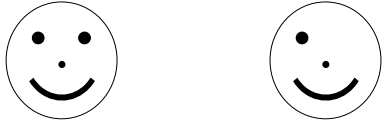} %left bottom right top
\caption[Resonant and dissonant images]{Examples of resonant and dissonant images.  The image on the left resonates with the internal model that generates a face.  The dissonant image lacks one eye and it is hard to recognise.  In other words, it is hard to use the image to make an inference.  The perceptual system acts to finesse inference, or gain resonance, which in this case would involve revealing the missing eye.} 
\label{fig:smiley}
\end{figure}

%Ordering and symmetry obsessions are the most primitive manifestation of OCD, and reveal the cause of the other types.
% Cite Summerfeldt1999 here on incompleteness
\vspace{-4mm}
Summerfeldt \cite{summerfeldt1999} cites an example of a 38-year-old man who had obsessions with themes of: `(1) the need to know or remember details, (2) the need for exactness in behavior and precision of expression, and (3) the need for symmetry and sameness in his physical environment (e.g., his appearance, the alignment of books, the condition of belongings). \dots Distress centered on not the content of obsessions, but \dots \, on a tormenting sense of hyperawareness and dissatisfaction.' 
\nind 
The explanations given earlier for checking, fear of harm and intrusive thoughts have implicated a dysfunction of the SMS.  The dysfunction was identified as a failure of inference  associated with a failure to obtain `resonance' in the distribution of $p(N|U)$ at sub-surface levels of the cognitive hierarchy. With ordering and symmetry obsessions, the involvement of the SMS is less obvious\footnote{Although some ordering obsessions are associated with a fear of harm if the associated ritual is suppressed\comment{need citation}.}, but the underlying dysfunction affects other functions that rely on modelling external events. The point is illustrated using Figure \ref{fig:smiley} which shows examples of resonant and dissonant images.  The image on the right is dissonant because it is close to a common form but it has one part missing, so it evokes an error at a sub-surface level of inference.  The situation is one where we can potentially finesse the inference by some action, such as moving our perspective to reveal the hidden eye.  Such an action is normal ordering behaviour which is aimed at bringing about recognition, or resonance. In OCD, this process occurs with ordinary scenes, leading to them appearing `not just right' \cite{coles2003}, and resulting in a compulsion to act.  It is the response to prediction error that explains not just ordering behaviour but the dysfunction of the SMS implicated in other manifestations of OCD.
%Ordering behaviour is seen in normal individuals and can considered as part of perception.  Some order in the environment is necessary for perception to occur and if the environment is disordered one option might be change it.  Examples of normal ordering behaviour are straightening a painting that is askew or tidying a desk.  Ordering implies changing the environment so that it better reflects internal models.   From this perspective it can be seen as an attempt to minimise entropy consistent with Karl Friston's free energy formulation of brain function \cite{fristonfree}.  Disorder implies unpredictability, so making an environment more ordered reduces the cognitive effort needed to interpret it.  
%____________________________________________________________________________________________________
%
%   F I G U R E
%____________________________________________________________________________________________________

\section{Discussion}
% Sum up.  Keep this brief
We have presented a hierarchical narrative framework of cognition and used it to propose an explanation of OCD.  The central concept, the formal narrative, follows from the use of generative models in cognitive inference. The framework consists of the following claims,
\vspace{-3mm}\paragraph{1)} A hierarchical model of narratives leads to the idea that the human perspective is composed of a hierarchy of layers, each of which contributes to the individual's perspective or `view of the world'. This idea is reconciled with a unitary perception by considering situations when we do not believe what we see, such as visual illusions.

\vspace{-3mm} \paragraph{2)} Prospective and retrospective narratives provide a model for mental time travel. Narratives are generated autonomously and constantly by the brain to model the past, present and future. 

\vspace{-3mm}\paragraph{3)} The security motivational system (SMS) generates a collection of narratives in response to a threat. Rather than passively identifying threats, the SMS actively tries to find evidence for the created scenarios.  This process is similar to a corporation performing penetration testing in which unlikely risk scenarios are created and evaluated. 

\vspace{-3mm}\paragraph{}
The idea that perception is generated is well-established, and it is just a small step to suggest that the generative process can operate autonomously.  In \cite{friston2011} Friston \etal proposed that `the same generative models used to make predictions can be deployed to predict the actions of self or others by simply changing the bias or precision (i.e. attention) afforded to proprioceptive signals.'.  A similar idea for employing generative models for prospection, retrospection and mental imagery was put forward in \cite{thesis}. 
\nind
We suggest that OCD arises from a dysfunction in sub-surface levels of inference while global inference is broadly unimpaired.  The dysfunction of inference is caused by an over-creative generative process, which leads to a unusually broad range of potential narratives.  The consequent lack of resonance is sensed as the external world being `not just right', and its automatic correction by the perceptual system is experienced as compulsion.  Ordering and symmetry obsessions are thus the effects of the perceptual system trying to achieve precise inference. The failure of inference affects the view not only of current scenes, but also of prospective and retrospective views which the SMS relies on.  The dysfunction leads to an individual not having a `clear view' of prospective scenes, leading to fear of harm or intrusive thoughts.  Action by the SMS to protect the individual and finesse the inference is felt as a compulsion to re-check.
\nind
Ian Jakes considered the qualities of a good explanation of OCD, citing Imre Lakatos \cite[p154]{jakes}, who identified research as \emph{theoretically progressive} if it leads to new predictions and \emph{empirically progressive} if the predictions are successful.  In this context prediction can be of a fact that is already known as well as predicting a new finding. One of the limitations of psychological explanations of OCD is that they are not formal, and do not make quantitative predictions.  The same limitation applies to the explanation in this paper in that the exact form of the hierarchical narrative model is yet to be defined.  However there is potential for a formal model which could then be applied to electrophysical data, so closing the explanatory gap between the neural and psychological explanations.

% Finish very briefly with
% 1. Limitations - this is not a formal model and only begins to close the explanatory gap
% 	The nature of the dysfunction is not precisely made
%	The model is not formal
%	Mention Ian Jakes' criteria - check old versions for this.
% 2. Proposals 
%	Generation of predictions of neuropsychological evidence. Predict new phenomena
% Then finish.

\newpage
\bibliographystyle{IEEEtran}
\bibliography{narr}
\end{document}

%% file: macpap2.tex
\newcommand{\be}{\begin{equation}}
\newcommand{\ee}{\end{equation}}
\newcommand{\bean}{\begin{eqnarray}}
\newcommand{\eean}{\end{eqnarray}}
\newcommand{\bea}{\begin{eqnarray*}}
\newcommand{\eea}{\end{eqnarray*}}
\newcommand{\bfig}{\begin{figure}}
\newcommand{\efig}{\end{figure}}
\newcommand{\ba}{\begin{array}}
\newcommand{\ea}{\end{array}}

\newcommand{\nind}{\newline\indent}

\newcommand{\comment}[1]{}

\newcommand{\etal}{\emph{et al. }}
\newcommand{\etalns}{\emph{et al.}}